\documentclass{article}
\usepackage{graphicx} 
\usepackage{geometry}
\usepackage[ruled, linesnumbered, lined]{algorithm2e}
\usepackage{amsmath, amsfonts}
\usepackage[numbers]{natbib}
\usepackage{authblk}
\usepackage{booktabs}

\global\long\def\State{\mathcal{S}} 
\global\long\def\Action{\mathcal{A}}

\global\long\def\Item{\mathcal I}
\global\long\def\History{\mathcal H}

\global\long\def\R{\mathbb R}
\global\long\def\dataset{\mathcal D}
\global\long\def\E{\mathbb E}

\global\long\def\Ind{ \boldsymbol{1} }

\global\long\def\argmin{\operatorname{argmin}\;}
\global\long\def\argmax{\operatorname{argmax}\;}

\global\long\def\HR{\operatorname{HR}}

\global\long\def\DCG{\operatorname{DCG}}

\global\long\def\rand{\text{rand}}
\global\long\def\all{\text{all}}

\global\long\def\ss{\boldsymbol{s}}

\global\long\def\zz{\boldsymbol{z}}
\global\long\def\pp{\boldsymbol{p}}

\begin{document}

\title{Integrating Offline Reinforcement Learning with Transformers for Sequential Recommendation}

\author[1,*]{Xumei Xi}
\author[2]{Yuke Zhao}
\author[2]{Quan Liu}
\author[2]{Liwen Ouyang}
\author[ *]{Yang Wu}

\affil[1]{School of Operations Research and Information Engineering, 
Cornell University}
\affil[2]{Bloomberg}
\affil[*]{Work done while at Bloomberg.}

\date{\texttt{ xx269@cornell.edu, \{yzhao826, qliu268, louyang11\}@bloomberg.net, yangwufl@gmail.com}}

\maketitle

\begin{abstract}
    We consider the problem of sequential recommendation, where the current recommendation is made based on past interactions. 
    This recommendation task requires efficient processing of the sequential data and aims to provide recommendations that maximize the long-term reward. 
    To this end, we train a farsighted recommender by using an offline RL algorithm with the policy network in our model architecture that has been initialized from a pre-trained transformer model. The pre-trained model leverages the superb ability of the transformer to process sequential information.
    Compared to prior works that rely on online interaction via simulation, we focus on implementing a fully offline RL framework that is able to converge in a fast and stable way. 
    Through extensive experiments on public datasets, we show that our method is robust across various recommendation regimes, including e-commerce and movie suggestions. 
    Compared to state-of-the-art supervised learning algorithms, our algorithm yields recommendations of higher quality, demonstrating the clear advantage of combining RL and transformers. 
\end{abstract}

\maketitle

\section{Introduction}

    Recommender systems encompass almost all facets of everyday life, from applications in the entertainment industry like movie
    % and music~\cite{chen2001music} 
    recommendations~\cite{devi2018movie} to 
    % news~\cite{karimi2018news} and 
    merchandise suggestions~\cite{schafer1999ecommerce}. 
    Especially with the emergence of big data, navigating the boundless sea of information without the guidance of a recommender can be challenging for customers.
    Therefore, designing a recommender that produces relevant and helpful suggestions tailored for a variety of user profiles is gaining more and more attention in recent years. From a business perspective, a good recommender should also adapt to a user's evolving tastes over time, in order to maintain customer engagement.

    Traditional latent factor based methods including matrix factorization~\cite{koren2009matrixfactorization, hidasi2012alstensor} aim to decompose the sparse user-item interaction matrix as a set of feature vectors representing users and items. 
    The more overlap one particular user has with an item in the feature space, the more likely it is that we recommend this item to the user. However, the static nature of matrix factorization makes it unsuitable for predicting recommendations in a sequential manner, since the method is oblivious to the chronological order of the interaction history. 
    More recently, viewing the recommendation process from a sequential perspective has been adopted by several  works~\cite{tang2018caser, kang2018sasrec}, since it takes into account session-based user profiles in a more systematic way compared to previous methods like matrix factorization. 
     Whether the model architecture is convolutional neural network~(CNN)~\cite{tang2018caser} or recurrent neural network~(RNN)~\cite{kang2018sasrec}, the methodologies are all considered supervised learning.
     One key element is still missing here: modeling the dynamics of user-item interaction. 
     More specifically, sequential supervised learning methods overlook the interactive and evolving nature of recommender systems. As a result, they make decisions that may be optimal in the short term but suboptimal in the long run.
    
    To tackle the aforementioned issue, we utilize the framework of reinforcement learning (RL), which has undergone remarkable development in the field of interactive games~\cite{schrittwieser2020mastering} and robotic control~\cite{srinivas2020curl}. The two main parts of RL are the agent and the environment. The agent is the decision maker, while the environment represents a potentially unknown set of rules applied to the agent. 
    The goal of the agent is to learn a decision rule, referred to as a policy, that maximizes the reward accumulated over time, by repeatedly interacting with the environment to collect data. 
    Unlike supervised learning, the RL agent learns by exploring the environment through trial and error. Another important distinction is that the current action of the RL agent influences not only the immediate results but also the long-term ones, as pointed out in~\cite{sutton2018rlintro}. Consequently, RL is perhaps a more appropriate framework than supervised learning for recommender systems, as it captures the interactive and evolving aspects. 
    There have been a growing number of works in the recommender community that deploy RL to provide sequential recommendations in different settings~\citep{liu2020drr, lei2019udqn, wu2020see-pt}. The focus of our paper is to integrate the algorithmic framework of offline RL with the transformer architecture. 
    % We refer the reader to a comprehensive survey~\cite{afsar2022rsrlSurvey} on this topic.

     In this paper, we model the recommendation process as a sequential decision making problem and utilize offline RL combined with transformers to obtain a recommendation strategy that achieves good long-term performance. 
     The offline RL framework makes sure that we only infer from previously collected data without the need to perform further online interaction. For interactive games, it is natural to learn online by trial and error since all interactions can be made in a controlled environment. 
     For recommender systems, however, online interactions can be risky and time-consuming. Revenue loss is also possible if the deployed policy is not properly trained. 
     On the other hand, building a realistic and reliable simulation environment from the offline data to mimic the real environment is highly non-trivial. As a result, an offline approach is more direct and trustworthy compared to online and simulation-based methods. 
     % The reason for choosing offline RL is that we only want to draw inferences from previously collected data and prohibit any further interaction with the environment. 
     % By doing so, we circumvent the need of creating a reliable simulation system to train the RL agent. Our offline approach is more direct and trustworthy since it only relies on the data log. 
     We list our main contributions as follows.
    \begin{itemize}
        \item We propose a fully offline RL-based recommendation framework, with an efficient actor-critic algorithm under the hood. The specific algorithm we use is Critic Regularized Regression (CRR)~\cite{wang2020crr} that incorporates value regularization in a computationally efficient way, to avoid bootstrapping error commonly encountered in offline RL. 
        \item We harness the capacity of transformers to provide our RL algorithm with high-quality, sequentially encoded item embeddings. The transformer model we use is the distilled GPT-2 (DistilGPT2)~\cite{sanh2019distilbert}, the smaller version of Generative Pre-trained Transformer 2 (GPT-2)~\cite{radford2019gpt-2}. To the best of our knowledge, we are the first to integrate offline RL with transformers in recommender systems.
        \item To better integrate RL and transformers to facilitate learning, we propose a two-stage training strategy. In the first stage, we train the policy via supervised learning on a simple next-item prediction task. In the second stage, we continue to train the policy~(actor) alongside the critic that estimates the Q function. The rationale of our two-stage strategy is two-fold. The first stage~(supervised learning) helps stabilize the training of RL to converge faster\footnote{Our experiment shows that RL training without the first stage will suffer from slow convergence or instability early on. The next-item prediction task is both simple and effective to boost convergence. 
        %We will also show later that supervised learning alone outperforms more complicated methods.
        }. The second stage (RL) improves the policy obtained from the first stage and alleviates overfitting. Through extensive experiments, we show that our method significantly outperforms  
        % state-of-the-art 
        supervised learning methods that use sequential modeling.
    \end{itemize}

    % \paragraph{Organization.} 
    % In Section~\ref{sec: related_work}, we provide a literature review of various approaches for recommender systems. 
    % In Section~\ref{sec: prelim}, we offer the reader the background information required to comprehend the algorithmic framework. Specifically, we introduce MDPs, offline RL, and Transformers.
    % In Section~\ref{sec: algo}, we formally introduce our proposed algorithm. 
    % In Section~\ref{sec: numerics}, we present numerical results obtained from testing our method on real-world datasets and compare them with other methods.
    % In Section~\ref{sec: discussion}, we discuss the impact of our work and potential directions for future research. 

\section{Related Work} 
\label{sec: related_work}

    In this section, we give an overview of various approaches proposed for recommender systems. 
    For the purpose of this paper, we categorize the methods into two groups based on whether the method uses RL or not. 
    Most non-RL approaches are supervised learning methods, where the task is to estimate unobserved interactions based on a sparse subset of observed interactions. 
    % User-item interactions can generally be represented as ratings that reflect user preferences or as binary variables that indicate user interest. 
    On the other hand, RL-based methods model the recommendation process as a Markov decision process~(MDP) and use various algorithms for inference. 
    
    \subsection{Non-RL Methods} 
    \label{subsec:lit_non-RL}
    
        % The problem of recommender systems is not a new topic in scientific research and numerous techniques have been introduced.
        We first discuss supervised learning methods.
        Content-based filtering~\cite{mooney2000content} matches the characteristics of the items with the user profile, which represents the user's interests. 
        Collaborative filtering methods recommend items based on the preferences of users who share similar tastes. Compared to content-based filtering, collaborative filtering relies on information about similar users as guidance, making the users implicitly collaborate with one another, whereas content-based filtering does not. 
        A popular and powerful example in collaborative filtering is matrix factorization~\cite{koren2009matrixfactorization, hidasi2012alstensor, wang2006similarity}. The sparsely observed user-item interaction matrix is decomposed as a set of feature vectors for users and items. Based on this decomposition, we estimate each unobserved interaction via the inner product of the corresponding user feature vector and the item feature vector. The implicit assumption of matrix factorization is that the user-item interaction satisfies a linear relationship. 
        Beyond linear interaction, we can use deep neural networks to allow for more complex relations, as demonstrated in~\cite{he2017neuralCF, hidasi2015sessionbased}.
        % Another example of supervised learning approaches is logistic regression~\cite{mcmahan2013logistic}. However, it has been observed that logistic regression may struggle to generalize to interactions that are not frequently observed.
        
        In addition to supervised learning, the application of contextual multi-armed bandits to recommender systems~\cite{li2010contextual, chapelle2011thompson, zeng2016timevarying} provides a new direction to explore, by considering the interactive nature of the system.
        To balance the trade-off between exploration and exploitation, the method of Thompson sampling is used in~\cite{chapelle2011thompson} and upper confidence bound is employed in~\cite{li2010contextual}.
        The work~\cite{zeng2016timevarying} proposes a dynamical context drift model on the reward function to tackle the problem of time-varying preference. 
        % Another work~\cite{wang2017bandit} combines low-rank matrix factorization and contextual multi-armed bandits to achieve the best of both worlds. 

    \subsection{RL-based Methods} 
    \label{subsec:lit_RL}
    
        Compared to the aforementioned methods, the benefit of adopting the RL perspective is two-fold. 
        Firstly, it incorporates an important dynamic that the current recommendation has an impact on future ones. For example, if a user likes one movie from a trilogy, the probability of them favoring the other movies in the trilogy increases. 
        Secondly, the RL agent is oriented toward maximizing the long-term reward, rather than the immediate one. This property makes RL more effective at keeping users engaged for a longer period. 
        Because of the dynamic nature of RL, we need to be aware of potential distribution shift during training, which is central to many ideas in algorithmic design. We will discuss more on this in Subsection~\ref{subsec:offline}.
        % Unlike supervised learning in which a static input-output relationship is assumed, RL takes on a dynamic perspective. 
        
        In terms of specific RL algorithms, they can be classified into two groups: model-based and model-free methods. 
        Model-based methods first predict the environment dynamics and then find the optimal policy, while model-free methods directly learn the optimal policy without modeling the environment. 
        One example of model-based methods applied to recommender systems is~\cite{shani2012mdpbased}.
        As the reconstruction of the dynamics could be costly, researchers also adopt model-free methods. 
        We can further divide the model-free methods into three main categories: value-based, policy-based, and actor-critic methods. 
        Value-based methods find the optimal value function and calculate the optimal policy based on the value function. 
        Policy-based methods directly learn the optimal policy, bypassing value function estimation. 
        Actor-critic methods optimize both the value function and the policy simultaneously. 
        For recommender systems, several papers~\cite{zhao2018deers, lei2019udqn, fu2021dhcrs, gao2022valuepenalized} use Deep Q-Network (DQN), a popular value-based method. 
        Deep Deterministic Policy Gradient (DDPG) is a powerful policy-based method that has been adopted by~\cite{wang2018srl-rnn, liu2020drr}.
        Actor-critic methods are present in~\cite{chen2022off,ge2021fcpo, xiao2021offline}. 
        We comment that over the past few years, there have been a growing number of works applying RL to recommender systems. 
        For a more comprehensive review, we refer the reader to a recent survey~\cite{afsar2022rsrlSurvey}.

\section{Preliminaries} 
\label{sec: prelim}

    %In this section, we provide some preliminaries that set the stage for later sections. Specifically, we discuss three topics: decision process formulation for recommender systems, the framework of offline RL, and transformer architecture. 
    %We first describe the decision process formulation for recommender systems in Subsection~\ref{subsec:MDP}. 
    %Then, we introduce offline RL and the associated challenges it presents in Subsection~\ref{subsec:offline}. In addition, we discuss the general framework of actor-critic methods for offline RL and the method we choose for our problem. 
    %Lastly, we give a brief introduction to the transformer architecture in Subsection~\ref{subsec:transformers}, which includes the advantages of the self-attention mechanism as well as its non-recurrent architecture. More importantly, we discuss the relevance of transformers to our work. 
    In this section, we provide some preliminaries which set the stage for later sections. We first describe the decision process formulation for recommender systems in Subsection~\ref{subsec:MDP}. 
    Then, we introduce the notion of offline RL and the associated challenges it presents in Subsection~\ref{subsec:offline}. Specifically, we discuss the general framework of actor-critic methods for offline RL and the method we choose. 
    Lastly, we discuss the transformer architecture in Subsection~\ref{subsec:transformers}.
    
    \subsection{Markov Decision Process (MDP) for Recommender Systems} 
    \label{subsec:MDP}
    % \subsubsection{MDP Formulation for}
        We formulate the MDP as a tuple $(\State, \Action, r, P, \mu_0, \gamma)$, where $\State$ is the state space, $\Action$ is the action space, $r$ is the immediate reward, $P$ represents the state transition, $\mu_0 \in \Delta(\State)$ is the initial state distribution and $\gamma$ is the discount factor. 
        In the context of recommender systems, we view the users as the environment providing feedback and the recommender as the agent that aims to find a good recommendation policy, by interacting with the users. In the sequel, we explain what each element of $(\State, \Action, r, P, \mu_0, \gamma)$ represents in our setting. 
        
        The action space $\Action$ for the agent corresponds to an item candidate pool of appropriate size and each action $a$ is an item recommended to the user. Since several RL algorithms struggle to scale to large and discrete item spaces, item embeddings~\cite{liu2020drr} are often adopted to convert the discrete item space to a continuous one. %\xxdelete{, making the implementation of policy gradient possible.} 
        Another way to handle large item space is to use hierarchical structures~\cite{chen2019tpgr, fu2021dhcrs}, which requires prior categorization of items. 
        The reward function $r: \State \times \Action \to \R$ represents the feedback provided by the user after receiving the recommendation. In practice, the reward can be a numerical rating or a binary variable representing click or purchase. For different datasets, the construction of rewards might differ. We defer the specifics of the reward function to Subsection~\ref{subsec:parameter}.
        
        It is less clear what the state should be. Ideally, the state should contain information such as user preference and interaction history, based on which the recommendation is made. There are many ways to design state representation. The algorithm from~\cite{lei2019udqn} models the state as the user feature vector obtained from matrix factorization. In the same spirit, the deep reinforcement learning based recommendation framework (DRR) in~\cite{liu2020drr} proposed three different structures, based on concatenating feature interactions between items, or between users and items. 
        % Apart from using features, the work~\cite{lei2020gcqn} leverages user-item bipartite graph as state representation. 
        Other works that take on a more sequential perspective directly model the state as the most recent items a certain user has interacted with~\cite{wu2020see-pt}. To further differentiate between the items with positive feedback (high rating, purchase) and negative feedback (low rating, skipped), one can simply include the complete feedback history~\cite{fu2021dhcrs}, categorize the items into two subgroups~\cite{zhao2018deers}, or only include positive items~\cite{ge2021fcpo}. The aforementioned works all use a fixed sequence length for user history, with padding if needed. A recent paper~\cite{antaris2021sar} proposes a novel sequence length adaptation. 
        In this paper, we model the state as a fixed length of the most recent positive items from the user's interaction history, due to its simplicity and effectiveness.
        
        Because our state only includes the positive items, the state transition happens if and only if a positive rating is received. 
        When the rating is negative, the state remains unchanged.
        Suppose the current state is a vector $\ss_t = (s^1_t, s^2_t, \dots, s^{\ell}_t)$ with $\ell \ge 1$ being the sequence length and each $s^i_t$ for all $i \in[\ell]$ corresponds to an item ranking from old to new. We have the following state transition rule conditioned on the feedback of the current action $a_t$:
        
        \begin{equation}
            \ss_{t+1} = 
            \begin{cases}
                    (s^2_t, \dots, s^{\ell}_t, a_t), & \text{if } r(\ss_t, a_t) > 0 \\
                    \ss_t, &\text{otherwise.}
            \end{cases} 
            \label{eq:state_transition}
        \end{equation}
        We note that the transition rule is deterministic conditioned on the immediate reward. Namely, $P(\ss_{t+1} \vert \ss_t) = 1$, following the formula~(\ref{eq:state_transition}).
    
        Lastly, the discount factor $\gamma \in (0,1)$ governs how much we care about future rewards. To see that, we define some functions related to the notion of value. 
        For a policy $\pi: \State \to \Delta(\Action)$, we define the value function $V^\pi: \State \to \R$ as
        \begin{equation}
            \label{eq:V_definition}
            V^\pi (\ss) = \mathbb{E}_{\pi} \left[ \sum_{t=0}^\infty \gamma^t r(\ss_t, a_t) \; \Big\vert  \; \ss_0 = \ss \right],
        \end{equation}
        which represents the expected cumulative reward the agent collects if the initial state is $\ss_0=\ss$ and the agent follows policy $\pi$. 
        The state-action value function (Q function) is similarly defined as
        \begin{equation}
            \label{eq:Q_definition}
            Q^\pi(\ss,a) = \mathbb{E}_{\pi} \left[  \sum_{t=0}^\infty \gamma^t r(\ss_t, a_t) \; \Big\vert  \; (\ss_0, a_0) = (\ss, a) \right],
        \end{equation}
        which is the expected cumulative reward when starting from the designated state-action pair $(\ss,a)$. The goal of RL is to find the optimal policy such that the value averaged over the initial state distribution $\mu_0$ is maximized: 
        \[
            \max_{\pi} \; J^\pi = \mathbb{E}_{\ss \sim \mu_0} \left[ V^\pi(\ss) \right].
        \]
        Directly from the definitions~(\ref{eq:V_definition}) and ~(\ref{eq:Q_definition}), we can see that the agent is more myopic when $\gamma$ is small. When $\gamma$ is close to one, the agent takes the future reward into account and the resulting optimal policy is more farsighted. 
        For future reference, we define the advantage function as
        \begin{equation}
            \label{eq:A_definition}
            A^\pi (\ss,a) = Q^\pi(\ss,a) - V^\pi (\ss),
        \end{equation}
        which measures the potential benefit we might get if we update the current policy $\pi$. When $A^\pi(\ss,a)>0$, it means the current policy is not optimal and the action $a$ should be preferred at state $\ss$. When $A^\pi(\ss,a) < 0$, it implies action $a$ is less preferable at state $\ss$.

    \subsection{Offline RL} 
    \label{subsec:offline}
    
        In the setting of offline RL, we get a previously collected dataset $\dataset = \{ ( s, a, s', r) \}$, and further interaction with the environment is prohibited. 
        When generating the dataset $\dataset$, we can assume the agent follows a behavior policy $\pi^\beta$. 
        % In some cases, the dataset is generated by a set of different policies that are potentially non-Markovian and non-stationary. 
        Since further exploration is not allowed, there is a limit to which the agent can improve upon the behavior policy. 
        Intuitively, if there exists a region in the environment that is not well explored during data collection, the agent trained on offline data is not expected to perform well inside that particular region. Therefore, the learning goal for offline RL is to output a policy that is \emph{optimal given the data}. We also point out that the presence of distribution shift in offline RL needs careful treatment to avoid overconfidence in less explored regions. 
        % At the very least, the output policy should outperform the behavior policy. 
        In the sequel, we first give a high-level review of actor-critic methods in offline RL. Afterward, we introduce the specific algorithm we adopt for the recommender system problem.

    \subsubsection{Actor-Critic Methods for Offline RL}
        % Model-free methods, as opposed to model-based ones, do not model the environment and instead learn the optimal policy via estimated values. 
        The actor-critic framework has two main components: the actor updates $\pi$ to improve the policy and the critic approximates $Q^\pi$ to evaluate the policy. 
        We parameterize the Q function and the policy as $Q_\theta(\cdot, \cdot)$ and $\pi_\phi(\cdot \vert \cdot)$. 
        Based on the offline dataset $\dataset$, we perform the Q update on $\theta$ and policy update on $\phi$:
        \begin{alignat}{2}
             \theta^{k+1} & \gets \underset{\theta} {\argmin} \textstyle \E_{(s,a,s') \sim \dataset} \Big[ \big(  r(s,a) + \gamma \E_{a' \sim \pi_{\phi_k}(\cdot \vert s') }[Q_{\theta^k} (s',a') ] 
              - Q_{\theta}(s,a)  \big)^2  \Big] 
            \label{eq:Q_update} \\
            \phi^{k+1} & \gets \underset{\phi} {\argmax} \E_{s \sim \dataset, a \sim \pi_{\phi}(\cdot \vert s)} \left[Q_{\theta^{k+1}} (s,a) \right].  
            \label{eq:policy_update}
        \end{alignat}
        The Q update~(\ref{eq:Q_update}) performs temporal difference (TD) learning and 
        the policy update~(\ref{eq:policy_update}) sets the policy greedily according to the estimated Q function.
        
        One of the fundamental difficulties of offline RL is the distribution shift. Namely, the state-action visitation measure induced by the behavior policy can be very different from that induced by the learned policy. As a result, the policy evaluation can be highly off-policy, which in turn affects policy improvement. If implemented naively, actor-critic methods often suffer from extrapolation error when evaluating out-of-distribution state-action pairs. 
        Researchers have proposed different ideas to address the distribution shift. 
        We will review some of the ideas here. Interested readers may refer to a survey~\cite{levine2020offline}. 
        Importance sampling~\cite{precup2000eligibility} corrects the bias in the estimation caused by distribution shift but requires good coverage of the behavior policy and suffers from high variance. It has been applied to recommender systems by~\cite{chen2022off}. The distribution shift can also be handled by adding constraints to the policy update. For example, Bootstrapping Error Accumulation Reduction~(BEAR) proposed in~\cite{kumar2019bear} constrains the current policy to be close to the behavior policy in terms of maximum mean discrepancy. Other works~\cite{siegel2020keep, nachum2017trust} consider regularizing the learned policy through KL divergence. For application in recommender systems, the paper~\cite{ge2021fcpo} adopts a trust region method using KL divergence.
        The idea of using uncertainty estimation of the Q function can regularize the target values and produce conservative estimators. Uncertainty can be estimated by taking the variance across value estimations of an ensemble of Q functions~\cite{wu2021uwac}. An example of using value penalized Q-learning in recommender systems can be found in~\cite{gao2022valuepenalized}. 
        In this paper, we adopt the algorithm called CRR from~\cite{wang2020crr}, which is a policy-regularized actor-critic method. The regularization is achieved by constraining the KL divergence between the learned policy and the behavior policy. We choose this method because of its simple yet effective update. As we will see shortly, the KL divergence constraint provides a computationally efficient modification to the original actor-critic update, which helps us save memory and time when working on gigantic datasets. At the same time, CRR pairs up well with transformers and converges fast, as demonstrated in Section~\ref{sec: numerics}.

    \subsubsection{Critic Regularized Regression (CRR)}
        \label{subsec:crr}

        We introduce a recently proposed offline RL algorithm CRR~\cite{wang2020crr} and explain the reasons for choosing CRR for our application. CRR modifies the policy update~(\ref{eq:policy_update}) in the standard actor-critic framework by
        value-filtered regression. Specifically, we update the policy parameter according to
        \begin{equation}
            \label{eq:crr_max}
            \underset{\phi}{\argmax} \E_{(s,a)\sim \dataset} \left[ f(Q_\theta, \pi_\phi, s,a ) \log \pi_{\phi} (a\vert s) \right],
        \end{equation}
        where the filter $f$ is a non-negative and scalar function that is monotonically increasing in the first argument $Q_\theta$. The filter decides what state-action pair we encourage the algorithm to update, preventing $\pi$ to use the actions that are not present in the dataset. 
        In~\cite{wang2020crr}, they consider two different filters: binary filter $f = \Ind[\hat A (s,a)>0]$ and exponential filter $f = \exp \left(\hat A (s,a)/ \beta \right)$.
        Here, $\Ind$ is the indicator function, $\beta>0$ is a hyperparameter controlling the filter strength, and $\hat A$ denotes the estimation of the advantage function defined in~(\ref{eq:A_definition}). 
        In our implementation, we estimate the advantage function by averaging over $m$ realizations of the current policy: 
        \begin{equation}
            \hat A_k (s, a) = Q_{\theta^k}(s,a) - \frac1m \sum_{j=1}^m Q_{\theta^k}(s, a^j), \quad \text{with } a^j \sim \pi_{\phi^k} (\cdot \vert s).
            \label{eq:advantage_estimate}
        \end{equation}
        From another perspective, the maximization program~(\ref{eq:crr_max}) is solving the maximum weighted likeliho103od, where the weight is given by the filter function $f$. 
        By using the filter, we put more weight on state-action pairs that have a larger advantage. As we discussed before, a large advantage indicates that the policy needs more emphasis on the particular state-action pair. Compared to behavior cloning ($f \equiv 1$), using a filter that varies depending on the advantage estimation is more flexible and less conservative. Rather than fitting the behavior policy, CRR can better distinguish between the good and bad actions taken by the behavior policy. Since for the recommender system problem, the dataset is often collected by carrying out a combination of different policies, it is imperative for the algorithm to distinguish the good actions from the bad ones. By using a filter that encourages good actions and penalizes bad ones, CRR achieves exactly what we need. 
        % Therefore, CRR can output a policy that outperforms the behavior policy.
        In our experiment, we choose the exponential filter since it is softer and smoother compared to the binary one. As a result, we do not run the risk of filtering out too many samples. Therefore, the policy update is 
        \begin{equation}
            \label{eq:crr_policy_update}
            \phi^{k+1} \gets \underset{\phi}{\argmax} \E_{\dataset} \left[ \exp \left( \hat A_{k} (s,a) / \beta \right) \log \pi_{\phi_k} (a\vert s)  \right]. 
        \end{equation}
        The Q update stays unchanged, i.e.~we still use formula~(\ref{eq:Q_update}). 
        There is also a connection to policy regularization in CRR.
        As pointed out by both~\cite{wang2020crr} and~\cite{nair2020awac}, we comment that the policy update~(\ref{eq:crr_policy_update}) is equivalent to optimizing the Lagrangian with the form $\mathcal{L}_\beta (\pi_{\phi^k}) =\E_{\dataset}  \left[ \hat A_{k} (s,a) \right] + \beta \left( \varepsilon - D_{KL}(\pi_{\phi^k}(\cdot\vert s) \; || \; \pi_\beta(\cdot\vert s))\right)$. 
        % \[
        % 	\mathcal{L}_\beta (\pi_{\phi^k}) =\E_{\dataset}  \left[ \hat A_{k} (s,a) \right] + \beta \left( \varepsilon - D_{KL}(\pi_{\phi^k}(\cdot\vert s) \; || \; \pi_\beta(\cdot\vert s))\right).
        % \]
        The learned policy is constricted to be close to the behavior policy in terms of KL divergence. 
        With this regularization, we avoid the potential over-extrapolation error often observed in offline RL. 
        From this perspective, the CRR update~(\ref{eq:crr_policy_update}) is essentially the closed-form solution of the KL-divergence regularized policy iteration, which explains its efficacy at addressing distribution shift.

    \subsection{Transformers} 
    \label{subsec:transformers}

        Introduced in~\cite{vaswani2017attention}, transformers initiated a major paradigm shift and have become the go-to model for processing sequential data in natural language processing (NLP) and computer vision (CV) tasks. 
        Before the emergence of transformers, recurrent neural networks (RNNs) such as  long short-term memory (LSTM)~\cite{Hochreiter1997lstm} and gated recurrent units (GRUs)~\cite{cho2014gru} were the common models for processing sequential data. The limitations of RNNs are two-fold.
        RNNs process data in sequential order and the previous information is stored in a state vector which is updated after reading the current token. As the sequence progresses, the vanishing gradient problem is exacerbated, rendering the model unable to scale to longer sequences and synthesize distant knowledge. On the other hand, RNNs are difficult to parallelize since the data must be processed one by one.
        The first limitation can be addressed by adding self-attention mechanisms, which allow all previous states to be accessed. With a learned weight that represents the relevance to a previous state, RNNs with attention can retrieve related but far-off information which previously was inaccessible. As a result, the attention mechanism improves the performance of RNNs by integrating information over longer horizons. However, the second limitation still persists for RNNs with attention: the sequential nature of RNNs prohibits the utilization of modern GPUs with powerful parallel computing capabilities. Transformers resolve the inefficiency of RNNs by getting rid of the recurrent structure altogether. The success of transformers shows that the self-attention mechanism works well all by itself without the recurrent architecture. Transformers can be parallelized during training, opening the door to pre-trained language models such as 
        % BERT (Bidirectional Encoder Representations from Transformers)~\cite{devlin2018bert} and 
        GPT (Generative Pre-trained Transformer)~\cite{radford2019gpt-2, brown2020gpt-3}. The pre-trained models have substantial generalization potential in the sense that they possess strong zero-shot and few-shot learning capacities after being trained on massive and diverse datasets like the Wikipedia Corpus, WebText, and Common Crawl.

        In our setting, we adopt the transformer architecture precisely because of its ability to synthesize data over long horizons (which is a consequence of self-attention) and the benefit of parallel computing (which is due to non-recurrent architecture). Unlike some NLP tasks that require sequence-to-sequence models, our goal is to predict a single item to recommend based on previous interactions. Hence, we adopt an encoder-only model which takes the user-item interaction history as input, embeds the history into a latent space, position encodes the embeddings and at last, passes them through a self-attention unit. 
        The particular transformer we use is  DistilGPT2~\cite{sanh2019distilbert}, which is a distilled version of the smallest GPT-2~\cite{radford2019gpt-2} with around 82 million trainable parameters. The idea of knowledge distillation~\cite{hinton2015distilling} is used to compress a large model, referred to as the teacher, to a more compact one, the student, that can reproduce the behavior of the teacher. By distilling the GPT-2, the resulting model has similar performance with less memory cost and faster computation.

        Our proposed method, which combines RL and transformers, is not the first attempt at this particular integration. 
        The recently introduced decision transformer~\cite{chen2021decision} views the problem of RL as a conditional sequence model and thereby harnesses the simplicity and scalability of the transformer architecture. Another recent work~\cite{parisotto2020stabilizing} modifies the architecture of transformers and significantly improves the stability and learning speed in the regime of RL. 
        Compared to our work, we comment that the aforementioned papers did not address the recommender systems as we do. 

\section{Proposed Approach} 
\label{sec: algo}
    In this section, we formally introduce our method that combines offline RL and transformers to tackle the problem of sequential recommendation. 
    The overall framework consists of two stages. 
     In the first stage, we only train the actor (policy agent)
     by performing supervised learning on a next-item prediction task, where the loss is computed via cross-entropy loss. 
     The initial training of the actor helps stabilize the RL algorithm used in the next stage to converge faster. 
    In the second stage, we run CRR introduced in Subsection~\ref{subsec:crr} to train the actor and the critic (value agent) simultaneously until convergence. 

    \subsection{Algorithm}
        
        We denote the user-item interaction history as $\History = \{(u,i,r)\}$, where $u$ is the user, $i$ is the item and $r$ is the feedback (e.g. rating, click, purchase). Let $\Item$ denote the full item space and $I$ denote its cardinality. Based on the history $\History$, we pre-process the data $\History$ to match the MDP formulation introduced in Subsection~\ref{subsec:MDP} and get $\dataset  = \{(\ss, a, \ss', r)\}$. Recall that the current state $\ss = (s^1, s^2, \dots, s^\ell)$ represents the most recent interaction history, where $\ell \ge 1$ is a hyperparameter for the fixed history length. The action $a$ is the next item recommended to the user. The next state $\ss'$ is transitioned according to~(\ref{eq:state_transition}). The reward $r$ is obtained by appropriate shifting of the initial feedback. Details on the formulation of the reward are deferred to Subsection~\ref{subsec:parameter}. 
        
\begin{figure}
            \centering
           \begin{tabular}{ccc}
                \includegraphics[width=.45\textwidth]{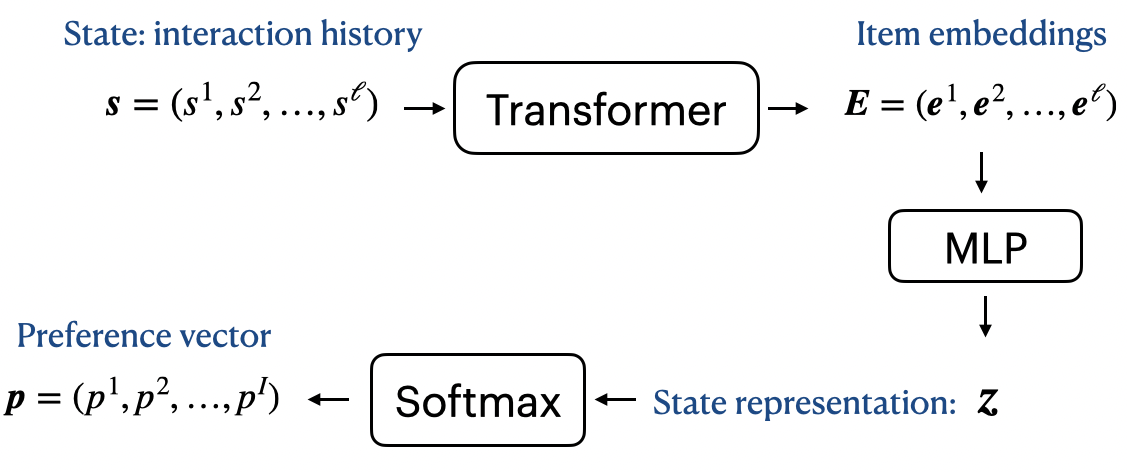} &  ~~~ & \includegraphics[width=.45\textwidth]{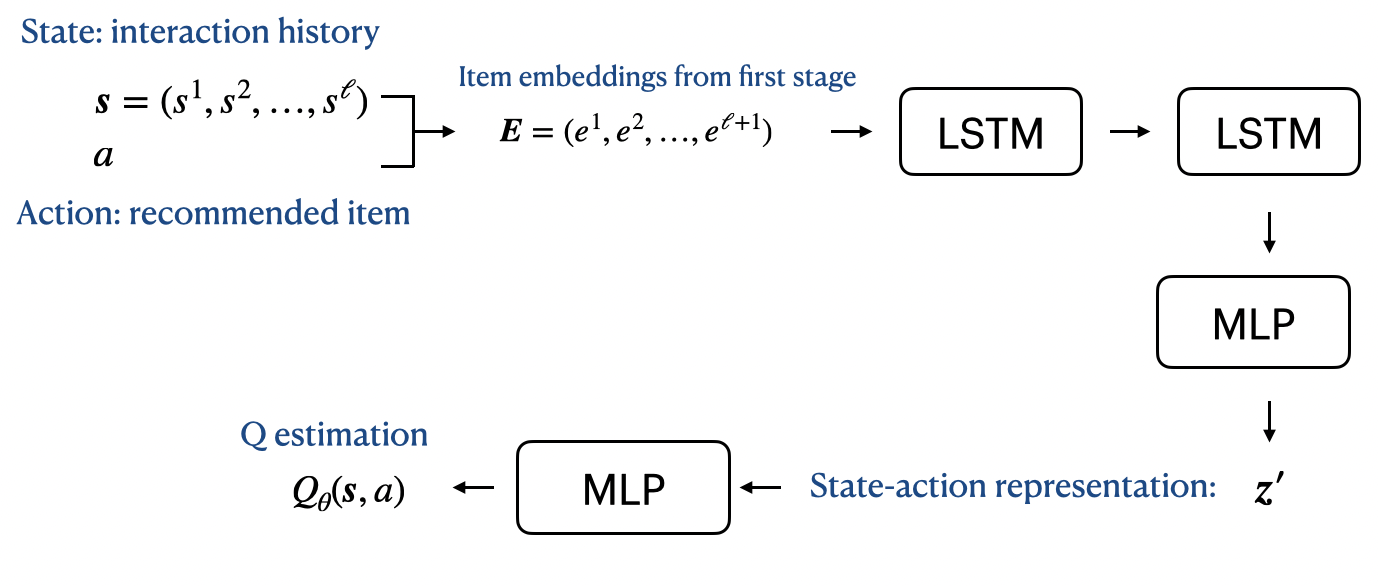} 
                \tabularnewline
           \end{tabular}
            \caption{Model architecture of the policy network~(left) and value network~(right). 
            For the policy network, the input is the state, which consists of the most recent interaction history of length $\ell$. The output is a probability distribution over the item space, which represents the probability of recommending each item to the user. 
            The item embeddings are vectors in $\R^d$ and the state representation $\zz$ is in $\R^I$. 
            MLP stands for multilayer perceptron, a fully connected feedforward network for extracting state representations. For the value network, the input is a state-action pair, and the output is an estimation of the Q function. Item embeddings are directly 
            from the training of the first stage. Two layers of LSTM are added before MLP. }
            \label{fig:actor_architecture}
        \end{figure}

        Our method includes two stages. The first stage uses supervised learning to train the policy on a next-item prediction task. 
        The architecture of the policy network is described in Figure~\ref{fig:actor_architecture}. 
        Specifically, we adopt DistilGPT2 as the transformer part because of its capability of processing sequential data as well as its computational advantage. 
        Although GPTs and DistilGPTs were initially designed for NLP tasks, their capability of processing sequential data makes their architectures suitable for much broader applications. For recommender systems, we treat each item as a word and a sequence of items with which a user interacted as a sentence. 
        In the first stage, the policy aims to mimic the behavior observed in the dataset by minimizing the cross-entropy loss defined below
        \begin{equation}
            H (\boldsymbol{p}(s), a)= - \sum_{i \in \Item} \Ind_{\{i = a \}} \log(p_i) = - \log(p_a).
        \end{equation}
        In the above equation, we are evaluating the cross-entropy between the output preference vector $\pp(s) \in \Delta(\Item)$ and the true one-hot vector $ \{ \Ind_{\{i = a \}} \}_{i\in \Item}$. 
        Since we want to encourage recommendations with positive feedback, we only use the data with $r>0$ for supervised training.
        % \xxedit{to boost convergence.}\yzcomment{"boost convergence" could be unclear with the previous "encourage recommendations with positive feedback".} 
        We denote the learned policy from this step as $\bar \pi$.
        The supervised learning step can be interpreted as a naive behavior cloning that only focuses on the immediate next recommendation. We have discovered that this is not enough to achieve long-term gain in the system, since the goal is to output the next-item prediction without thinking about the future. 
        
        Therefore, we construct a second stage in which we run offline RL with the policy~(actor) initialized exactly as the learned policy  $\bar \pi$ from the previous stage. The value network for the critic is similar to the actor, except that the transformer is replaced by LSTM\footnote{Performance is similar for LSTM and transformer as the critic. }. The full architecture is plotted in Figure~\ref{fig:actor_architecture}.  
        % \yzcomment{Is it better if we add a graph to show LSTM architecture? When I wrote the later section e.g. we set 2 layers for LSTM, I felt unclear.} \locomment{agree having a full architecture layout is better} 
        For computational efficiency, we extract the item embeddings from the training of the first stage and freeze them during the second stage as we discovered updating the embeddings did not help convergence. 
        %\yzcomment{Rewrite the following sentences, also add r=0 clarification.} 
        The first stage of supervised learning benefits the second stage by providing a reasonable starting point for RL. A good initialization not only stabilizes the algorithmic trajectory but also improves learning efficiency. The second stage, in turn, helps improve the myopic policy $\bar \pi$ trained from the first stage, since the objective of RL is to maximize the cumulative reward aggregated over future steps. %\xxcomment{I don't feel it's necessary to mention this. For a usual RL training, there is no reason to delete the data with non-positive rewards, to begin with. We deleted them probably because of some deficiency in the algorithm itself. It would make more sense to mention ``reincluding'' $r=0$ if you have an experiment without them.  Also, it is not clear what $r=0$ refers to in this section. We haven't introduced the datasets and how the rewards are designed. Readers might get confused. I strongly recommend we don't mention it here.} 
        %\xxdelete{It is worth noticing that to avoid a single reward setting and to further facilitate recommendations toward long-term gains, we include data with $r=0$ for RL training, which is different from supervised learning training. }
        We summarize our method in Algorithm~\ref{algo:crr_transformer}. We will discuss more on the choice of hyperparameters $b, \beta, \gamma$ and $\tau$ in Subsection~\ref{subsec:parameter}.
        
        \begin{algorithm}
            \caption{CRR-Transformer} \label{algo:crr_transformer}
            \KwData{$\dataset = \{(s,a,s',r)\}$. }
            \KwResult{Recommendation policy $\hat \pi$.}
            \emph{First stage (supervised learning)} - train initial policy using next-item prediction: 
            \[
                \bar \pi=\text{argmin}_\pi \E_{(s,a,r) \sim \dataset} \left[ H (\pi(s), a) \; \vert \; r > 0 \right].
            \] \\
            \emph{Second stage (RL)} - run CRR with policy initialization $\pi_{\phi} \gets \bar \pi$. Randomly initialize critic $Q_{\theta}$. Initialize target networks as: $\pi_{\phi'} \gets \pi_{\phi}, Q_{\theta'} \gets Q_{\theta}$. \\
            % Initialize critic network $Q_{\theta}$. Copy parameters of policy and critic networks to corresponding target networks $\pi_{\phi'}$, $Q_{\theta'}$. \\
            \For{$k=0$ \KwTo $K$}{
                Sample $\{(s_j, a_j, s'_j,r_j)\}_{j=1}^b$ from $\dataset$. \\
                Calculate advantage estimator $\hat A $ according to~(\ref{eq:advantage_estimate}). \\
                Update actor $\phi$ with gradient 
                \[
                     - \nabla_{\phi} \frac1b \sum_j \log \pi_{\phi} (a_j \vert s_j) \exp (\hat A (s_j, a_j ) / \beta).
                \]
                Update critic $\theta$ with gradient
                \[
                    \nabla_{\theta} \frac1b \sum_j \left[ Q_{\theta}(s_j, a_j) - ( r_j + \gamma \E_{a \sim \pi_{\phi'}(s'_j)}Q_{\theta'} (s_j', a) ) \right]^2.
                \]
                % Update policy with gradient: $\nabla_{\phi} -\frac{1}{b}\sum_i  \log \pi_{\phi}(a^i \vert s^i) f(Q_{\theta},\pi_\phi,s^i,a^i)$, where $ f = \exp(\hat{A}_\theta(s,a)/\beta)$, $\hat{A} = Q_{\theta}(s,a)-\frac{1}{m}\sum_j Q_\theta(s,a^j)$ with $a^j \sim \pi(\cdot \vert s)$\\
                % Update critic with gradient: $\nabla_{\theta}-\frac{1}{b}\sum_i [Q_{\theta}(s^i,a^i)-(r^i+\gamma \mathop{\mathbb{E}}_{a \text{\~} \pi_{\phi'}(s'^{i})}Q_{\theta'}(s'^{i},a))]^2$ \\
                Soft update target $\phi'$ and $\theta'$ with parameter $\tau$:
                \begin{align*}
                    \phi' &\gets \tau \phi + (1-\tau) \phi' \\
                    \theta' &\gets \tau \theta + (1-\tau) \theta'.
                \end{align*}
            } 
            Output final estimate $\hat \pi$.
        \end{algorithm}

\section{Experiments} \label{sec: numerics}

    In this section, we report the numerical results from implementing our algorithm CRR-Transformer. We conduct experiments on two public datasets and compare our method with other %\xxdelete{state-of-the-art} 
    supervised learning algorithms. 
    
    \subsection{Datasets}
        We conduct our experiments on MovieLens 25M~\cite{harper2015movielens} and Yoochoose~\cite{yoochoose}. 
        The MovieLens 25M (ML-25M) dataset contains 25 million 5-star movie ratings from 1995 to 2019. Each entry of rating records includes the user ID, movie ID, rating (0.5--5.0), and timestamp. It also contains features of the movies, computed via a machine learning algorithm based on user-contributed content such as tags, ratings, and textual reviews. We will not use the features in the dataset. Instead, we compute the features via the next-item prediction in the first stage of our algorithm. We argue that our approach preserves sequential information, which benefits learning in the second stage. 
        Additionally, the tag genome involves tag relevance scores for movies. As described in~\cite{vig2012ml}, the tag genome was calculated via a machine learning algorithm based on user-contributed content such as tags, ratings, and textual reviews. 
        Consequently, the tag genome encodes the level of certain properties a movie exhibits, such as atmospheric, thought-provoking, realistic, etc. 
        Hence, we can directly use the tag relevance scores as item embeddings. However, we discover that the item embeddings output by our supervised learning step is much superior to the tag relevance scores. For ML-25M, our training goal is to recommend movies for which the user gives high ratings.
        
        The Yoochoose dataset contains six months of user activities for a large e-commerce company that sells a variety of merchandise such as clothes, toys, electronics, etc. The dataset includes both the clicks and purchases performed by the users. Each entry, representing a click or purchase event, contains the item ID, item price, quantity, session ID, and timestamp. 
        % We keep both click and purchase datasets for Yoochoose, in which each entry represents a click or purchase event. Each entry contains the item ID, item price, quantity, session ID, and timestamp. 
        Since our model only focuses on recommending one item regardless of the price, we ignore the quantity and price information in the implementation. For Yoochoose, we aim to recommend items the user will click on and even purchase. We will see shortly how the rewards are constructed for click and purchase. 
        We provide some statistics of the two datasets in Table~\ref{tab:dataset}.

        \begin{table}[h]
            \centering
            \begin{tabular}{c | c | c}
                 Dataset & \# of records & \# of items   \\
                 \hline
                 ML-25M & 25,000,095 &  62,423 \\
                 Yoochoose & 1,177,769 & 52,739
            \end{tabular}
            \caption{Dataset statistics.}
            \label{tab:dataset}
        \end{table}

        In the ML-25M dataset, it is allowed to have multiple records for the same user at the same time. However, this clearly violates our model assumption under which only one item is recommended at a time. To mitigate this issue, we delete the duplicate entries at the same timestamp and only keep the single records. This conflict between the datasets and our modeling can be mostly attributed to the fact that the timestamps are not sufficiently fine-grained. For the ML-25M dataset, it is impossible for a person to watch and review two movies at the same time. There must be a sequential order that is not reflected by the current timestamp. We argue that a random shuffling of simultaneous records would also work. We delete the overlapping entries for efficiency and simplicity.

        In terms of train-test split, there are two major ways. One is a chronological split and the other is a random split by users. We discover that the first chronological way typically leads to slightly worse performance, compared to the second one. This phenomenon can be explained by different degrees of distribution shift.
        When split based on users, the distribution shift from the training set to the test set depends on how similar the users behave. If the users are sampled in an i.i.d.~way, then there is no distribution shift.
        On the other hand, splitting chronologically forces us to do out-of-time prediction, which is difficult due to the potential shift of user preference over time. 
        We decide to split the training set and test set according to the timestamp, as this is the most natural way from the sequential perspective: the training is done in the past and we try to predict recommendations in the future. 
        Specifically, for ML-25M, the earliest 99\% of the data is used during training. The next 0.2\% in chronological order is the validation set and the latest 0.8\% is the test set. 
        For Yoochoose, the training stage uses the early 98\% of data. The next 0.2\% is used for validation and the latest 1.8\% is for testing.
        %Specifically, for ML-25M, the early 99\% of the data is used during training and the later 1\% is left for validation and testing. 
        %For Yoochoose, the training stage uses the early 98\% of data, and the later 2\% is used for validation and testing.

    \subsection{Evaluation Metrics} 
    \label{subsec:metrics}
        
        We discuss how to evaluate and compare different methods. We consider four metrics in total, two of which are based on hit rate (HR) and the rest on normalized discounted cumulative gain (NDCG).
        
        HR is calculated by comparing the next recommendations with the true answer. Since the true answer only exists when the reward is positive, we average over all positive reward entries. Let $n^+$ denote the number of positive reward entries in the validation or test set, namely $n^+ = \sum_{j \in[n]} \Ind_{\{r_j>0\}}$. From a candidate pool of items, a prediction model selects the top $10$ items with the highest scores, denoted by $ \Item_j = \{ i_j^1, \cdots, i_j^{10} \}$ for some data entry $j $. 
        The average HR@10 over data entries that have positive rewards is defined as
        \begin{equation}
             \HR @ 10 = \frac{1}{n^+} \sum_{j: r_j > 0} \Ind_{\{a_j \in \Item_j \}} ,
            \label{eq:hr10}
        \end{equation}
        where $a_j$ is the true item liked by the user ($r_j > 0$). 
        For each sample, the quality of the recommendation is determined by whether the true item is in the top $10$ recommendations. The position of the true item in the recommendation ranking does not matter. 
        On the other hand, NDCG takes into account the ranking of recommended items and therefore provides a more fine-grained view of the recommender performance~\cite{jarvelin2002cumulated}. To understand how it works, we first define the discounted cumulative gain in which the relevance of each item is weighted based on its position
        \[
            \DCG @ 10 = \frac{1}{n^+} \sum_{j : r_j>0} \sum_{h=1}^{10} \frac{\Ind_{\{ a_j = i_j^h \}} }{\log_2(h+1) }.
        \]
        The higher the rank of the true item is, the larger DCG is. 
        However, the magnitude of DCG is dependent on the number of items recommended. To resolve this issue, we can normalize DCG by the ideal DCG score when we recommend the most relevant items first. Because there is only one relevant item in our case, IDCG is binary and therefore NDCG coincides with DCG.

        For each sample $j$, we construct two different action spaces: $\Action_j^{\rand}$ and $\Action_j^{\all}$. 
        To decrease the space cardinality and thereby reduce the difficulty of estimation, quite a few prior works~\cite{lei2019udqn, wu2020see-pt, antaris2021sar} selected the action space $\Action_j^{\rand}$ to be the union of the true answer $\{a_j\}$ and a fixed number of randomly selected items unrated up till this timestamp. In this paper, we set the number of randomly selected items to be $100$.
        We refer to the HR@10 and NDCG@10 calculated by using $\Action_j^{\rand}$ as HR@10-rand and NDCG@10-rand, respectively. However, such metrics sometimes cannot truthfully reflect the performance of a recommender system as the values can rely on the choices of random items in the action space, especially when the size of the item universe is much larger than the size of the random items. 
        We believe a more accurate metric should consider all unrated items to form the action space. Thus, we define $\Action_j^{\all}$ as the candidate pool containing all the previously unrated items. We use HR@10 and NDCG@10 to denote the metrics obtained from using $\Action_j^{\all}$. 
        Because $\Action_j^{\all}$ can be significantly larger than $\Action_j^{\rand}$, HR@10 and NDCG@10 are usually lower than HR@10-rand and NDCG@10-rand, and also much harder to improve. 

    \subsection{Baselines} 
    \label{subsec:baseline}
        Before we compare Algorithm~\ref{algo:crr_transformer} with other methods, we demonstrate the first stage (supervised learning of policy) alone is already superior to several state-of-the-art supervised learning methods for sequential recommendation. We list the baseline methods we compare against as follows.
        % We compare two state-of-the-art sequential recommendation models with distilGPT2 (denoted as Transformer) for supervised learning to illustrate distilGPT2's efficacy in generative sequential prediction. Based on this, we are able to apply offline RL algorithm with the pre-trained transformer instead of baselines as the initial actor to yield higher quality results.

        \begin{itemize}
        	\item Caser~\cite{tang2018caser} is a CNN-based approach to capture the order of interactions. A sequence of recent items is embedded as an ``image'' in the time and latent spaces. Sequential patterns can thus be learned with convolutional filters.
         
        	\item SASRec~\cite{kang2018sasrec} is a self-attention based sequential model to allow both long-term semantics and short-term preferences. It utilizes a point-wise feed-forward network on top of the attention mechanism to incorporate non-linearity and to consider interactions between different latent dimensions.
        \end{itemize}

        The learned policy from the first stage of Algorithm~\ref{algo:crr_transformer} is only an intermediate result, but is already better than both Caser and SASRec (see Figure~\ref{fig:learning_curve_ml} and~\ref{fig:learning_curve_yc}). This gain is precisely due to the transformer's remarkable ability to process sequential data. As a result, we argue that combining RL with the transformer architecture is far better than with other architectures such as CNN or RNN. In the sequel, we refer to the first stage of Algorithm~\ref{algo:crr_transformer} as ``Transformer-only'', to distinguish it from the full two-stage method.

    \subsection{Experimental Setup} 
    \label{subsec:parameter}  
           
        We first discuss the values of the hyperparameters. 
        For both datasets, the state sequence vector contains the most recent $30$ items. If the sequence length is less than $30$, we pad the sequence from the beginning. We use the Adam optimizer for all optimizations. The learning rate is chosen from $\{10^{-4}, 10^{-3}, 5 \times 10^{-3}\}$. The dropout ratio is chosen from $\{0, 0.05, 0.1, 0.2\}$. For the CRR algorithm in the RL stage, we use the exponential filter with $\beta=1$ and clip the filter value with an upper bound of $20$ for stability. The value network (critic) has two layers of LSTM with hidden size of $1,024$. For advantage estimation~(\ref{eq:advantage_estimate}), we sample $m=4$ actions and take the average\footnote{There are other ways of advantage estimation. In our case, averaging works the best. We choose a small $m$ for efficiency reasons. }. Both the policy parameter and the Q parameter are updated in a soft manner with $\tau=0.01$.
        The discount factor $\gamma$ is set as $0.6$ for ML-25M and $0.9$ for Yoochoose, respectively. 
        % The minibatch size is set as $b=128$. 
        For more details, please refer to Appedix~\ref{appen:more}.

        In terms of the immediate reward, we construct simple discrete reward functions that effectively encourages convergence and stabilizes training. For ML-25M, to encourage recommendations of relevant movies that receive high ratings from the user, we give a positive reward for movies with ratings greater than or equal to a threshold of $3.5$:
        \begin{equation}
            r_{\text{ML}}  = 
                \begin{cases}
                    1 & \text{if rating} \ge 3.5 \\
                    0 & \text{otherwise}. \\
                \end{cases}
            \label{eq:reward_ML}
        \end{equation}
        For Yoochoose, to promote purchase behaviors, we set a higher reward for purchase events. Specifically,  we have
        \begin{equation}
            r_{\text{YC}}  =
                \begin{cases}
                    3 & \text{if purchase} \\
                    1 & \text{if click}.
                \end{cases}
            \label{eq:reward_YC}
        \end{equation}

        In the first stage (supervised learning), we only train the policy and evaluate the model performance on the validation set at the end of each epoch. We then select the model at the epoch that achieves the highest validation score (HR@10) and we pass the policy parameters on to the second stage as its initialization. During the second stage, we evaluate the performance on the validation set every $1,000$ batches and output the policy with the highest validation score (HR@10). 
        For ML-25M, the evaluation takes the next true movie with a positive reward as the target. For Yoochoose, the evaluation takes the next true item, whether clicked or purchased, as the target.

    \subsection{Results} 
    \label{subsec:results}
        In this subsection, we present and discuss the results. 
        
        \subsubsection{Learning Curves}
        \label{subsubsec:learning_curves}
            \begin{figure*}
                \centering
                \includegraphics[width=0.8\textwidth]{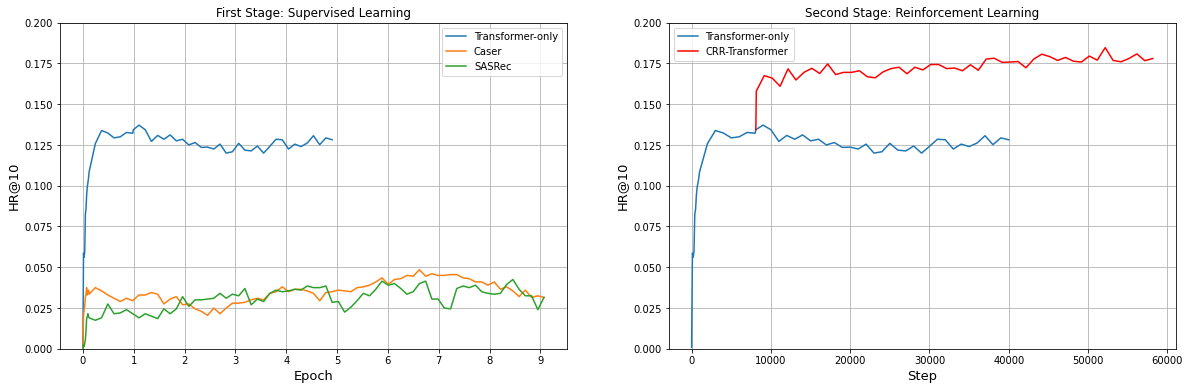}
                \caption{Learning curves for the first stage (supervised learning) and second stage (RL) on ML-25M dataset. The evaluation metric is the HR@10 score averaged over $3$ experiments. For the second stage, the policy is initialized from the first stage checkpoint with the best validation score.}
                \label{fig:learning_curve_ml}
            \end{figure*}
    
            \begin{figure*}
                \centering
                \includegraphics[width=.8\textwidth]{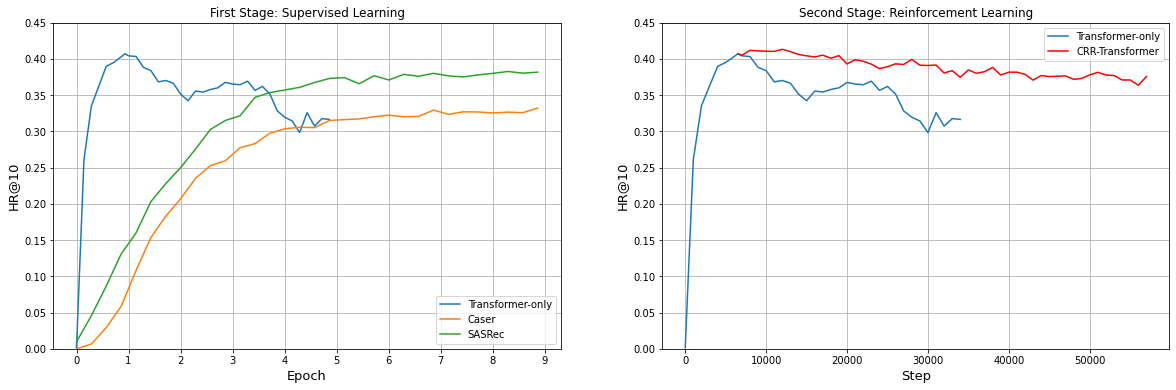}
                \caption{Learning curves for the first stage (supervised learning) and second stage (RL) on the Yoochoose dataset. The evaluation metric is the HR@10 score averaged over $3$ experiments. For the second stage, the policy is initialized from the first stage checkpoint with the best validation score.}
                \label{fig:learning_curve_yc}
            \end{figure*}

            Figure~\ref{fig:learning_curve_ml} and~\ref{fig:learning_curve_yc} present the learning curves of the two stages in our CRR-Transformer approach on validation data for ML-25M and Yoochoose, respectively. We observe that for the first stage (supervised learning ), Transformer-only reaches the best evaluation performance quickly. Even though it suffers from overfitting later, the peak of Transformer-only's HR@10 score still outperforms baselines across all epochs, which justifies our choice of Transformer-only as the initialization seed for the second stage. At the second stage, the offline RL algorithm continues to improve the HR@10 score.

        \subsubsection{Test Results}
        \label{subsubsec:test_results}

            \begin{table*}
                \caption{Test set results on HR@10, HR@10-rand, NDCG@10, NDCG@10-rand with the ML-25M dataset. We show the scores averaged over $3$ experiments as well as the standard deviations. Boldface highlights the highest average score. }
                \label{tbl:ml}
                \footnotesize
                \centering
                \begin{tabular}{lccccr}
                    \toprule
                     & Caser & SASRec & Transformer-only & CRR-only & CRR-Transformer\\
                     \midrule
                    HR@10  & $0.0275 \pm 0.0020$ & $0.0270 \pm 0.0012$ &  $0.1303 \pm 0.0146$ &  $0.1627 \pm 0.0035$ & \pmb{$0.1840 \pm 0.0010$}  \\
                    HR@10-rand & $0.3815 \pm 0.0083$ & $0.4115 \pm 0.0030$ & \pmb{$0.8876 \pm 0.0069$} &  $0.8689 \pm 0.0042$ & $0.8737 \pm 0.0026$  \\
                    NDCG@10 & $0.0128 \pm 0.0005$ & $0.0131 \pm 0.0007$  & $0.0676 \pm 0.0078$  & $0.0889 \pm 0.0022$ & \pmb{$0.1019 \pm 0.0008$} \\
                    NDCG@10-rand &  $0.2632 \pm 0.0064$ & $0.2794 \pm 0.0005$ &  $0.6502 \pm 0.0213$ & $0.6672 \pm 0.0063$
                    & \pmb{$0.6850 \pm 0.0022$}  \\
 
                \bottomrule
                \end{tabular}
            \end{table*}
        
            % \begin{table*}
            %     \caption{Test set results on HR@10, HR@10-rand, NDCG@10, NDCG@10-rand with the ML-25M dataset. We show the scores averaged over $3$ experiments as well as the standard deviations. Boldface highlights the highest average score. }\label{tbl:ml}
            %     \footnotesize
            %     % \hspace*{-1.0cm}
            %     \centering
            %     \begin{tabular}{lcccr}
            %     \toprule
            %     Models & {HR@10} & {HR@10-rand} & {NDCG@10} & {NDCG@10-rand}  \\
            %     \midrule
            %     Caser  & $0.0275 \pm 0.0020$  & $0.3815 \pm 0.0083$  & $0.0128 \pm 0.0005$  & $0.2632 \pm 0.0064$   \\ 
            %     SASRec  & $0.0270 \pm 0.0012$  & $0.4115 \pm 0.0030$  & $0.0131 \pm 0.0007$  & $0.2794 \pm 0.0005$    \\ 
            %     Transformer-only  & $0.1303 \pm 0.0146$  & \pmb{$0.8876 \pm 0.0069$}  & $0.0676 \pm 0.0078$  & $0.6502 \pm 0.0213$  \\ 
            %     CRR-Transformer  & \pmb{$0.1840 \pm 0.0010$}  & $0.8737 \pm 0.0026$  & \pmb{$0.1019 \pm 0.0008$}  & \pmb{$0.6850 \pm 0.0022$}  \\ 
            %     \bottomrule
            %     \end{tabular}
            % \end{table*}

            \begin{table*}
                \caption{Test set results on HR@10, HR@10-rand, NDCG@10, NDCG@10-rand with the Yoochoose dataset. We show the scores averaged over $3$ experiments as well as the standard deviations. Boldface highlights the highest average score. }
                \label{tbl:yc}
                \footnotesize
                \centering
                \begin{tabular}{llccccr}
                    \toprule
                    & & Caser & SASRec & Transformer-only & CRR-only & CRR-Transformer\\
                     \midrule
                    Purchase & HR@10  &  $0.1733 \pm 0.0036$  & $0.1948 \pm 0.0051$ &  $0.2274 \pm 0.0110$ & $0.2424 \pm 0.0110$ & \pmb{$0.2559 \pm 0.0065$} \\
                    & HR@10-rand & $0.7659 \pm 0.0117$  & $0.7202 \pm 0.0130$  & $0.8216 \pm 0.0093$ & $0.7479 \pm 0.0099$ & \pmb{$0.8253 \pm 0.0061$}   \\
                    & NDCG@10 & $0.0906 \pm 0.0015$  & $0.1083 \pm 0.0027$  &  $0.1199 \pm 0.0059$ & $0.1288 \pm 0.0055$  & \pmb{$0.1374 \pm 0.0041$}  \\
                    & NDCG@10-rand &  $0.6333 \pm 0.0133$ &   $0.6003 \pm 0.0081$ &  $0.7114 \pm 0.0108$ & $0.6455 \pm 0.0106$  & \pmb{$0.7237 \pm 0.0032$}   \\
                    \midrule 
                    Click & HR@10 & $0.3236 \pm 0.0035$ & $0.3801 \pm 0.0015$ & $0.3792 \pm 0.0054$ & $0.3386 \pm 0.0052$ &  \pmb{$0.3973 \pm 0.0063$}  \\
                    & HR@10-rand &  $0.8258 \pm 0.0020$ &   $0.8117 \pm 0.0084$ & $0.8553 \pm 0.0081$ & $0.8065 \pm 0.0050$ &  \pmb{$0.8580 \pm 0.0006$}  \\
                    & NDCG@10 & $0.2013 \pm 0.0017$  &   $0.2350 \pm 0.0024$ & $0.2371 \pm 0.0068$  & $0.2034 \pm 0.0025$ &  \pmb{$0.2541 \pm 0.0047$}  \\
                    & NDCG@10-rand & $0.7252 \pm 0.0046$ &  $0.7291 \pm 0.0056 $ & $0.7715 \pm 0.0074$ &  $0.7127 \pm 0.0034$ & \pmb{$0.7775 \pm 0.0021$} \\ 
                \bottomrule
                \end{tabular}
            \end{table*}

            % \begin{table*}
            %     % \caption{Test set results on HR@10, HR@10-rand, NDCG@10, NDCG@10-rand with the Yoochoose dataset. We show the scores averaged over $3$ experiments as well as the standard deviations. Boldface highlights the highest average score. }
            %     % \label{tbl:yc}
            %     \footnotesize
            %      \hspace*{-1.5cm}
            %     \centering
            %     \begin{tabular}{lcccrcccr}
            %     \toprule
            %     Models
            %          & \multicolumn{4}{c}{purchase}
            %          & \multicolumn{4}{c}{click}
            %     \\
            %      \cmidrule(lr){2-5} 
            %      \cmidrule(lr){6-9} 
            %      & {HR@10} & {HR@10-rand} & {NDCG@10} & {NDCG@10-rand}  
            %      & {HR@10} & {HR@10-rand} & {NDCG@10} & {NDCG@10-rand}  
            %     \\
            %     \midrule
            %     Caser  & $0.1733 \pm 0.0036$  & $0.7659 \pm 0.0117$  & $0.0906 \pm 0.0015$  & $0.6333 \pm 0.0133$  & $0.3236 \pm 0.0035$  & $0.8258 \pm 0.0020$ & $0.2013 \pm 0.0017$ & $0.7252 \pm 0.0046$  \\ 
            %     SASRec  & $0.1948 \pm 0.0051$  & $0.7202 \pm 0.0130$  & $0.1083 \pm 0.0027$  & $0.6003 \pm 0.0081$  & $0.3801 \pm 0.0015$  & $0.8117 \pm 0.0084$ & $0.2350 \pm 0.0024$ &  $0.7291 \pm 0.0056 $\\ 
            %     Transformer-only  & $0.2274 \pm 0.0110$ & $0.8216 \pm 0.0093$ & $0.1199 \pm 0.0059$ & $0.7114 \pm 0.0108$  & $0.3792 \pm 0.0054$ & $0.8553 \pm 0.0081$ & $0.2371 \pm 0.0068$ & $0.7715 \pm 0.0074$  \\ 
            %     CRR-Transformer  & \pmb{$0.2559 \pm 0.0065$}  & \pmb{$0.8253 \pm 0.0061$}  & \pmb{$0.1374 \pm 0.0041$}  & \pmb{$0.7237 \pm 0.0032$} &  \pmb{$0.3973 \pm 0.0063$}  & \pmb{$0.8580 \pm 0.0006$}  & \pmb{$0.2541 \pm 0.0047$}  & \pmb{$0.7775 \pm 0.0021$}  \\ 
            %     \bottomrule
            %     \end{tabular}
            % \end{table*}   

            Table~\ref{tbl:ml} and~\ref{tbl:yc} summarizes the performance of all methods in terms of the four evaluation metrics for the ML-25M and Yoochoose datasets, respectively. 
            CRR-only refers to running CRR without pre-trained transformer initialization, i.e.~we skip the first stage of Algorithm~\ref{algo:crr_transformer} and start the second stage with random initialization for both the policy and the value networks. For a fair comparison, we stop CRR-only at the same iteration as CRR-Transformer~(first stage and second stage combined).
            We observe that CRR-only underperforms CRR-Transformer in both datasets. This could be caused by the slow convergence or instability of RL training, which again demonstrates the necessity to initialize the RL training with a pre-trained transformer that has effectively learned the sequential information.
            
            In Table~\ref{tbl:ml} that shows the test results for ML-25M, we can see that Transformer-only already outperforms both baselines across all scores, which demonstrates the effectiveness of DistilGPT2 to model sequential information. Based on the high-quality policy output by Transformer-only, our CRR-Transformer approach is able to further improve upon it by taking long-term interactions into account and differentiating between relevant and irrelevant movies via the binary reward signal~(\ref{eq:reward_ML}).  
            The RL stage improves all scores except for HR@10-rand. 
            Since the action space $\Action^{\all}$ can be significantly larger than $\Action^{\rand}$, predicting with $\Action^{\rand}$ is much easier than with $\Action^{\all}$. As a result, HR@10-rand may not accurately reflect the quality of the recommendation. Moreover, CRR-Transformer achieves a higher NDCG@10-rand average score than Transformer-only. This shows that even with the same randomly selected universe, CRR-Transformer is more likely to recommend relevant movies at top positions. Furthermore, CRR-Transformer also results in much smaller standard deviations than Transformer-only for all metrics, confirming the stability and robustness of our two-stage training process.

            % Table~\ref{tbl:ml} summarizes the top-10 performance on recommendations for good-quality movies with the ML-25M dataset. In this case, we regard movies with $rating \geq 3.5$ as good-quality movies. We can see that the Transformer-only model has already beaten both baselines across all metric scores on average, which demonstrates the effectiveness of distilGPT2 to model sequential information. Based on this, the proposed CRR-Transformer approach is able to take long-term interactions into account and further differentiate between good-quality movies and bad-quality movies with the binary reward signal. The RL stage promotes all the average metric scores except for HR@10-rand. Considering that the action space is large, HR@10-rand may not be an accurate measurement of performance. Besides, CRR-Transformer achieves higher NDCG@10-rand average scores than Transformer, showing that even with the same randomly selected universe, CRR-Transformer is more likely to recommend good-quality movies at the top positions. CRR-Transformer also results in much smaller standard deviations than Transformer for all metrics, which verifies the stability and robustness of our training process.

            For the Yoochoose dataset, the performance of all methods on both the purchase and click tasks is shown in Table~\ref{tbl:yc}. Different from the learning curves, we split the test set evaluation by event types (click or purchase). Transformer-only achieves better average scores than the two baselines excluding HR@10 for the click task, where SASRec performs slightly better over Transformer-only. Nevertheless, CRR-Transformer achieves the highest average scores consistently across all metrics. We argue that during the RL stage, the model learns to focus more on purchase events that have higher rewards. We can see that CRR-Transformer outperforms other methods by a larger margin on the purchase task, which aligns with our recommendation goal, and also verifies that the agent successfully learns from the additional long-term rewards.

            In summary, the consistent improvement in both datasets demonstrates the effectiveness and robustness of our approach across different recommendation regimes and goals. 

        \subsubsection{Sensitivity to Discount Factor}
        \label{subsubsec:sensitivity}
            \begin{figure*}
                \centering
                \includegraphics[width=0.8\textwidth]{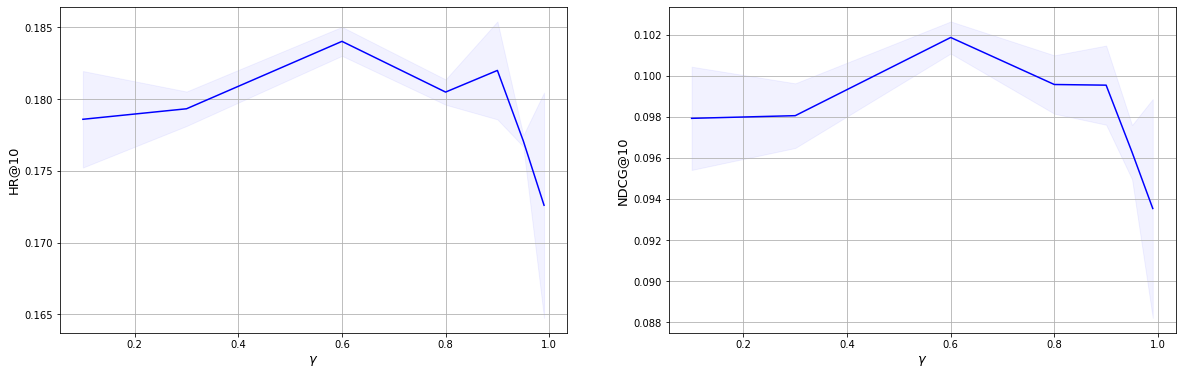}
                \caption{Sensitivity to the discount factor for ML-25M. The line and band refer to the average and standard deviation of HR@10 and NDCG@10 over $3$ experiments. The discount factor $\gamma$ is chosen from $\{ 0.1, 0.3, 0.6, 0.8, 0.9, 0.95\}$. }
                \label{fig:sensitivity_ml}
            \end{figure*}
    
            \begin{figure*}
                \centering
                \includegraphics[width=.8\textwidth]{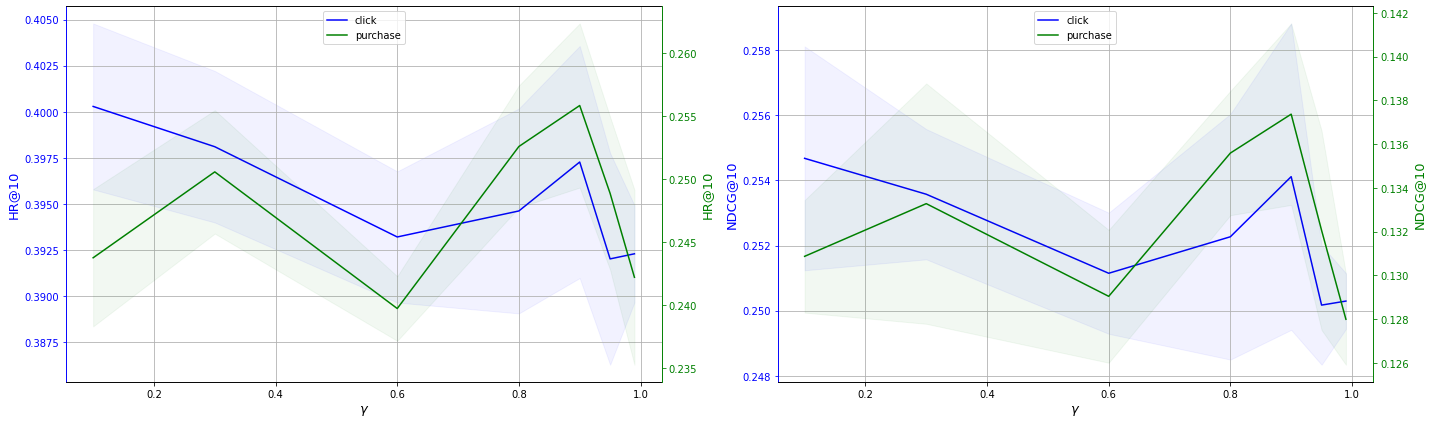}
                \caption{Sensitivity to the discount factor for Yoochoose. The blue line and band refer to the average and standard deviation of HR@10 and NDCG@10 for the click task over $3$ experiments. The green line and band refer to the average and standard deviation of HR@10 and NDCG@10 for the purchase task over $3$ experiments. The discount factor $\gamma$ is chosen from $\{ 0.1, 0.3, 0.6, 0.8, 0.9, 0.95\}$. }
                \label{fig:sensitivity_yc}
            \end{figure*}

            Finally, we discuss the sensitivity of our results to the discount factor $\gamma$. We focus on the test results of CRR-Transformer under different values of $\gamma$. Figure~\ref{fig:sensitivity_ml} and~\ref{fig:sensitivity_yc} illustrate the sensitivity of HR@10 and NDCG@10 to $\gamma$ for the ML-25M and Yoochoose datasets, respectively. The tight range of metric scores shows that HR@10 and NDCG@10 are not overly sensitive to the choice of $\gamma$. Furthermore, there are certain trends we can observe. 
            For ML-25M, both the average HR@10 and NDCG@10 improve when $\gamma$ gradually increases from $0.1$ to $0.6$. 
            However, the performance saturates around $0.6$ and starts to drop as $\gamma$ continues to increase. Not only does $\gamma=0.6$ yields the highest average score, it also results in the smallest standard deviation, indicating that the algorithm is more stable at $\gamma=0.6$. 
            % Figure~\ref{fig:sensitivity_ml} and Figure~\ref{fig:sensitivity_yc} illustrate the sensitivity of CRR-Transformer results to discount factor $\gamma$ with ML-25M and Yoochoose datasets respectively. Clearly for ML-25M dataset, both the average HR@10 and NDCG@10 improves when the discount factor $\gamma$ increases from a smaller number 0.1 to 0.6. They reach their maximums at $\gamma=0.6$ and then start to drop. Also, the smaller standard deviations of metric scores at $\gamma=0.6$ indicates that the algorithm is relatively stable with this discount factor. 
            For Yoochoose, the click task and the purchase task show different sensitivity to the discount factor. 
            The purchase task has a similar sensitivity trend as the ML-25M case: the score first increases and then decreases. The only difference is that it achieves peak performance at $\gamma=0.9$. The click task, on the other hand, gets better results when $\gamma$ is relatively small. A small $\gamma$ indicates that the model emphasizes more on the immediate reward rather than the long-term gains. Since the purchase events usually involve longer interactions between the user and items, compared to click events, this change in sensitivity trend indeed makes sense: to achieve a successful purchase, the agent needs to be incentivized with future rewards. 

            % For Yoochoose dataset, click and purchase results shows different sensitivity trends to discount factor. The purchase recommendation has similar sensitivity trend as the ML-25M case, except that it achieves the best performance at $\gamma=0.9$. The click recommendation, on the other hand, achieves better result when $\gamma$ is very small. Small $\gamma$ indicates that the model emphasizes more on the immediate reward but less on the long-term gains. Considering that purchase events usually involve more interactions between the user and items than click events, we argue that the sensitivity trends make sense.  

\section{Discussion} \label{sec: discussion}
    We propose a fully offline RL algorithm for recommender systems, leveraging the sequential benefit provided by the transformer architecture. 
    Our results demonstrate that our method is superior to several supervised learning baselines, which confirms that RL combined with transformers is a powerful tool.
    By combining offline RL and transformers, our model not only inherits the sequential information learned by the transformers but also attains substantially better long-term performance. 
    However, there are still limitations to the current approach: KL divergence style regularization might be overly pessimistic when it comes to out-of-distribution actions. 
    For future work, we plan to find the balance between over-estimation (caused by optimism) and under-estimation (caused by pessimism). 
    To that end, we can improve the offline RL algorithm by incorporating ideas from both online and offline RL literature. 
    For instance, we can borrow ideas from proximal policy optimization for the policy updates. Furthermore, we can explore ways to incorporate uncertainty into value updates. Another exciting direction is to integrate RL with transformers in a more systematic way. 
    %For future work, we plan to investigate the use of different RL algorithms with transformers. Furthermore, we can explore other ways to integrate RL and transformers to further improve performance. %\locomment{is it a good direction? I suppose we picked CRR out of other RL frameworks for some reasons. Maybe talk about uncertainty as a future direction? @quan @xumei, any thoughts?}

\appendix 

\section{More on experimental setup}
\label{appen:more}

    In this appendix, we provide more details on the hyperparameters. 
    For ML-25M, the learning rates for Transformer-only, Caser, and SASRec are $0.0001$, $0.0001$, $0.001$, respectively. The batch size is set as $128$.
    The item embedding size is $50$ for the baselines and $768$ for DistilGPT2 as default. For Caser, the model consists of two vertical convolution filters and two horizontal filters with heights chosen from $\{1, 2, 3, 4, 5\}$. The dropout ratio is $0.05$. For SASRec, following the original paper~\cite{kang2018sasrec}, we set the self-attention as single-head. The number of self-attention blocks is set as $2$. The item embedding size is $50$. The drop-out ratio is $0$. 
    For Yoochoose, the learning rates for Transformer-only, Caser, and SASRec are $0.001$, $0.001$, $0.005$, respectively. The batch size is set as $128$ for DistilGPT2 and $256$ for the baselines. The item embedding size is $64$ for the baselines and $768$ for DistilGPT2 as default. For DistilGPT2, the residual dropout is set as $0.2$. For Caser, we use one vertical convolution filter and $16$ horizontal filters with heights chosen from $\{2, 3, 4\}$. The dropout ratio is $0.1$. For SASRec, both the numbers of heads and self-attention blocks are set as $1$. The drop-out ratio is $0.1$. 
    %For CRR, the learning rate is set as $0.0001$. The batch size is set as $128$. The dropout ratio is set as $0.2$. 
    For the CRR algorithm in the RL stage for both datasets, the learning rate is set as $0.0001$. We also use a cosine annealing scheduler with no learning rate warmup. The minibatch size is set as $128$. The dropout ratio is set as $0.2$. 

\newpage
\bibliographystyle{plain} 
\bibliography{reference}

\end{document}